\documentclass[jkps,preprint,fleqn,showkeys]{revtex4}

\usepackage[pdftex]{graphicx}
\usepackage{amssymb}
\usepackage{amsmath}
\usepackage{bm}

\usepackage{multirow}

\begin{document}
\setcounter{page}{0}

\title[]{Fragment productions in DJBUU and SQMD: comparative study}

\author{Dae Ik \surname{Kim}}
\email{di.kim.phys@gmail.com}
\thanks{Current address: Gyeonghae girl's high school, Jinju 52651, Korea}
\affiliation{Department of Physics, Pusan National University, Busan 46241, Korea}
\author{Kyungil Kim}
\affiliation{Rare Isotope Science Project, Institute for Basic Science, Daejeon 34000, Korea}

\author{Chang-Hwan Lee}
\email{clee@pusan.ac.kr}
\affiliation{Department of Physics, Pusan National University, Busan 46241, Korea}

\author{Youngman Kim }
\affiliation{Rare Isotope Science Project, Institute for Basic Science, Daejeon 34000, Korea}

\author{Sangyong Jeon}
\affiliation{Department of Physics, McGill University, Montreal, Quebec, Canada H3A2T8}

\date[]{Received 24 June 2022}

\begin{abstract}
We study $^{208}$Pb+$^{40,48}$Ca reactions at  $E_{\rm beam}$ = 50, 100 AMeV with DJBUU and SQMD transport codes. We compare  the large primary fragments from the two codes at the end of the simulation time. We observe that overall the two models produce similar fragments. However, we see a noticeable difference between DJBUU and SQMD at $E_{\rm beam}=100$ AMeV with the impact parameter $b$ = 0 fm and discuss this difference in terms of the difference in the equation of state adopted in the two models and the difference in stability  inherent to  the BUU-type and  QMD-type models.
\end{abstract}


\maketitle


\section{Introduction}

Worldwide, there are many new rare isotope production facilities coming online in the foreseeable future.
One of the new facilities under construction is RAON (Rare isotope Accelerator complex for ON-line
experiments) by the Rare Isotope Science Project(RISP) of Institute for Basic Science (IBS) in
Korea~\cite{Tshoo:2013voa}. RAON will provide an ample amount of exotic neutron- and proton-rich isotopes
and enrich various advanced research fields: nuclear physics, astrophysics, and applications in medical,
material, and biological science. For a brief summary of RAON physics, see~\cite{Jeong:2018qvi}.

The evolution of rare isotope (RI) science evolves through interplay between three basic
ingredients: experiment, simulation and theory. To make a major progress in RI science,  we must
make use of all the three approaches. 
A good example is the nuclear symmetry energy.
The nuclear symmetry energy originates from the
difference in the proton and the neutron numbers in dense nuclear matter or nuclei and it is an important
physical quantity in understanding the properties of finite nuclei such as the neutron-skin  thickness of
heavy nuclei, neutron star properties and the collapse of the core of a massive star. The nuclear symmetry
energy may also play an important role in modeling the r-process nucleosynthesis.
To study the symmetry energy in a terrestrial environment laboratory, we perform heavy
ion collision experiments at RI facilities such as RAON.
To connect experimentally observed quantities such as neutron
to proton ratios are then compared with nuclear transport model calculations.
The two transport models we are using in this study, DaeJeon
Boltzmann-Uehling-Uhlenbeck (DJBUU) and Sindong Quantum Molecular Dynamics (SQMD),
are general purpose models for low-to-intermediate energy heavy ion collisions but 
the primary thrust for developing them has been to support physics outcome at RAON.

In this work we compare features of BUU- and QMD-type transport models in general and specifically DJBUU
and SQMD by studying fragment productions from the both models.  For this we study fragment productions in
$^{208}$Pb+$^{40}$Ca and $^{208}$Pb+$^{48}$Ca reactions at $50$ and $100$ AMeV with the impact parameters
$b=0,~3,~6$ fm. Since the most direct comparison of different transport models is to compare primary
fragments before using an afterburner in the simulations,  we focus on the primary fragments from DJBUU
and SQMD. For fragment production simulations in heavy ion reactions, we refer to~\cite{Mocko:2008rj,
Gaitanos:2007mm, Napolitani:2009pc, Liu:2014poa}.


\section{Model descriptions}

Transport theory describes the dynamics of nuclear collisions at Fermi/intermediate
energies semi-classically.
The key quantity is the phase space density (coordinate and momentum distribution). Transport
simulations in general take the following steps: initialization of the projectile and target nuclei,
propagation of the (test) particles, hadron-hadron scatterings with Pauli blocking.

There are two complimentary transport model approaches. The first one is the BUU-type models whose goal is
to calculate the one-body phase space distribution. 
Due to the test-particle ansatz and the mean-field potential, 
the BUU-type model is deterministic and contains no fluctuations except the numerical ones. 
The second one is the QMD-type models that are based on an N-body Hamiltonian.
As the QMD models do not use the test particle method, they 
are better suited for event-by-event analysis of heavy ion collisions.

In the next section,
we will briefly describe the DJBUU model and the SQMD model. For more details on these models we refer
to~\cite{DJBUU2016, Kim:2020sjy} for DJBUU and \cite{Kim:2017hmt} for SQMD. 
For a recent review of most of the transport models that are
currently being used in low-to-medium energy heavy ion collisions,
see~\cite{TMEP:2022xjg}.

\subsection{DJBUU model}

DJBUU mostly uses widely accepted standard methods in transport theory.
One distinctive feature is the profile
function of the test particle
\begin{equation}
g({\bf u}) = g(u) = {\cal N}_{m,n} (1 - (u/a)^m)^n
\ \ \ \hbox{for}\  0 < u/a < 1
\label{eq:profile_func}
\end{equation}
where $u = |{\bf u}|$ is either the displacement vector or the relative momentum,
${\cal N}_{m,n}$ is the normalization constant, and $m > 1$ and $n > 1$ are positive integers. 
The parameter $a$ controls the width of the function and for $|u/a| > 1$, $g(u) = 0$. 
Unlike the usual gaussian profile function,
$g(u)$ vanishes smoothly at $u=a$ and it is exactly integrable.
The default values are $m=2$, $n=3$ and $a_x = 4.2\,\hbox{fm}$ 
for the position profile and $a_p = s\hbar/a_x$ for the momentum profile with $s \approx 0.6$.

Another feature of the DJBUU model is the mean field potential. In addition to a widely-used Walecka type
relativistic mean field potential, we use a relativistic mean field potential obtained from an extended
parity doublet model~ \cite{Motohiro:2015taa, Shin:2018axs,Takeda:2017mrm}
to study the origin of the baryon mass and partial chiral symmetry
restoration in dense nuclear matter.
In this model, the scalar field is the chiral partner of the pion field. 

Now, we describe how we identify the produced fragments in DJBUU. Since the BUU-type models use the test
particle method, it is in general not easy to define a nucleon cluster as a single isotope. We remark
here that to identify a light cluster whose mass number is smaller than four, the composition production
method was employed in the collision terms ~\cite{Danielewicz:1992mi}.

Within the energy range in which we are interested, 
a large cluster will form near the center of a collision.
As such, we identify the cluster as follows in DJBUU simulations.
During our simulations, we span three-dimensional space of $-$80.0 fm $<$ $x$, $y$, $z$ $<$ 80.0 fm in the
center of mass frame by dividing the space into $160\times 160\times 160$ unit cells of 1\,$\hbox{fm}^3$
volume.
Then, we search for all cells that has the the baryon density equal to or larger than $0.1\,\rho_0$
where the nuclear saturation density $\rho_0 = 0.17$ fm$^{-3}$.
Once we find all such cells, we regard the collection of these cells as a cluster
or a fragment 
and count all the test particles for protons and neutrons in the cluster. 
The total number of the test particles is then divided by
the number of the test particle used in our simulation (100 in the current study) to assign the
proton and neutron numbers of the cluster. The value of the proton and neutron of a cluster 
is rounded up to the nearest integer.
We should mention here that
since DJBUU and also other transport models do not include nuclear pairing effects,
which are essential for the odd-even staggering in nuclei,
there is an inherent uncertainty in assigning the proton and neutron numbers
to the cluster.

\subsection{SQMD model}

In SQMD  a standard gaussian type of the nucleon wave function is used,
\begin{align}
\psi_i(\vec{r},t) = \frac{1}{(2\pi \sigma_r^2)^{3/4}}\exp \left(-\frac{(\vec{r}-\vec{r_i})^2}{4\sigma_r^2} + \frac{i}{\hbar}(\vec{p_i}\cdot \vec{r})\right)\,,
\end {align}
where $r_i$ and $p_i$ are the position and the momentum of the $i$th nucleon center, respectively. $\sigma_r$ is the width of a single nucleon in the coordinate space. We use $\sigma_r=1.3$ fm to ensure the stability of the initialized nuclei.

The interaction among nucleons is described by the Skyrme potential which has been widely used in the QMD models such as \cite{Aichelin:1991xy,Papa:2002xqa}. The Hamiltonian is given by
\begin{align}
H &= \sum_i \frac{\vec p_i^2}{2m_i} + U_{\rm tot}, \label{Hamiltonian} \\
U_{\rm tot} &= U_{\rm Skyrme} + U_{\rm surf} + U_{\rm sym} + U_{\rm coul}\, ,
\end{align}
where $m_i$ is the mass of the $i$-th nucleon. The density dependent Skyrme interaction term is given by
\begin{align}
U_{\rm Skyrme} = \frac{\alpha}{2} \left(\frac{\rho}{\rho_0} \right) + \frac{\beta}{\gamma + 1} \left(\frac{\rho}{\rho_0}\right)^\gamma \; ,
\end{align}
where $\rho$ is the baryon number density.
We take  $\alpha=-218$ MeV, $\beta=164$ MeV and $\gamma=4/3$. With these parameters we obtain the incompressibility  $K$=236 MeV and the binding energy $-16$ MeV at the saturation density.
For initialization, propagation and collision, SQMD adopts the standard methods used in transport models 
\cite{Kim:2017hmt}.

In SQMD simulations,
we identify clusters using the Minimum Spanning Tree (MST) algorithm. 
After a collision,
we measure the distances among all the nucleons and connect two nucleons with the minimum distance to span
all nucleons with the shortest distance. If the minimum distance of a nucleon pair is longer
than $3.5$ fm, then the two nucleons are not connected. We identify the connected lumps as
primary fragments.


\section{Results}

In this section, we compare the ``biggest fragments (BFs)" in DJBUU and SQMD simulations. Here, the BF
means the fragment that has the largest mass number in each  simulated event and the BFs means the
aggregation of the BF from each simulation. As described in the previous section, DJBUU and SQMD use
different methods to define the fragments.

We recall that DJBUU uses the test-particle method. In this method,
each simulated event contains an ensemble of events.
When calculating physical quantities such as baryon number densities, we effectively
average over the ensemble in each time step. 
The BF in DJBUU means the largest cluster formed near the center of a collision at the end
of the simulation time, and the BFs in our study is the collection of the BF from ten runs. While, the BF
in SQMD is the fragment which has the largest mass number among fragments at the end of the simulation
time, and the BFs are the aggregation of such a BF from ten-thousand runs.

We simulate  $^{208}$Pb+$^{40}$Ca and $^{208}$Pb+$^{48}$Ca reactions at $50$ and $100$ AMeV with the
impact parameters $b=0,3,6$ fm.
Both in DJBUU and SQMD, the initialized projectile and target are placed
$16$ fm apart from each other. Here, 16 fm is the distance between the center of the projectile and
target. Then, we do the simulations until $t=300$ fm/c. In the case of DJBUU, we use one hundred test
particles for each nucleon  and perform ten runs. We perform ten thousand runs for the SQMD simulations
and also check some of our results with twenty-thousand runs to confirm that the results are robust
against the number of runs. Since we are interested only in the central region we plot up to $20$ fm 
in each of the transverse directions.

\begin{figure} 
\includegraphics[width=\linewidth]{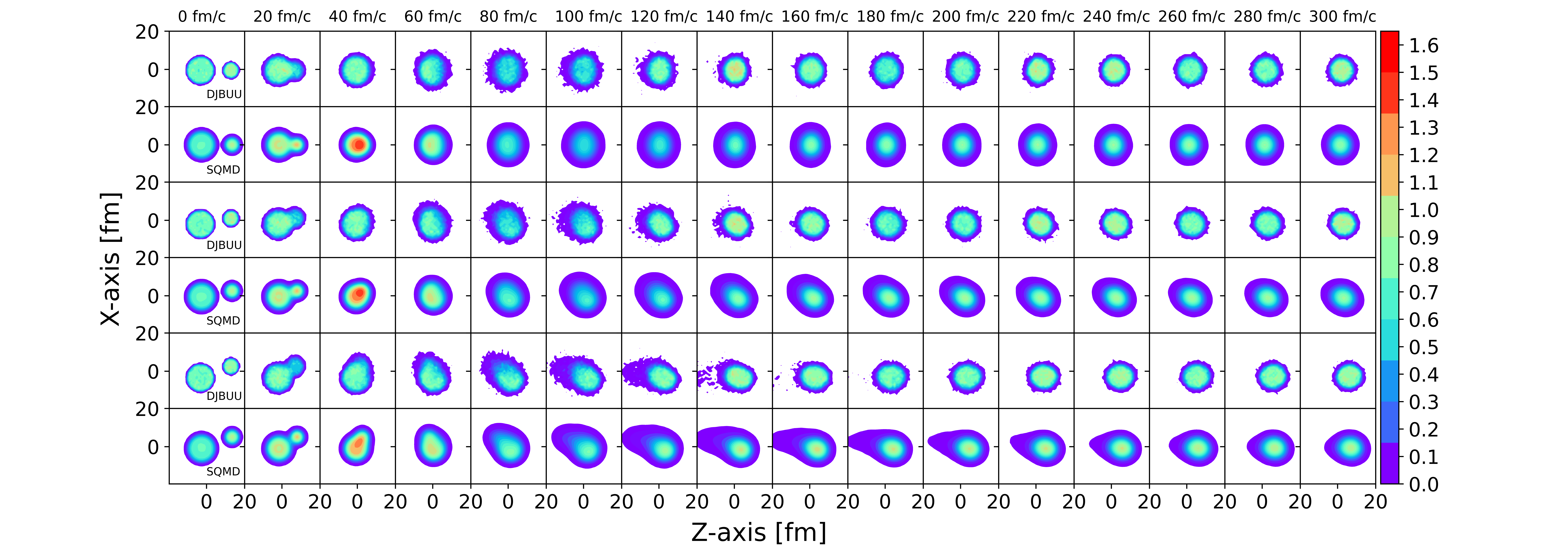}
\includegraphics[width=\linewidth]{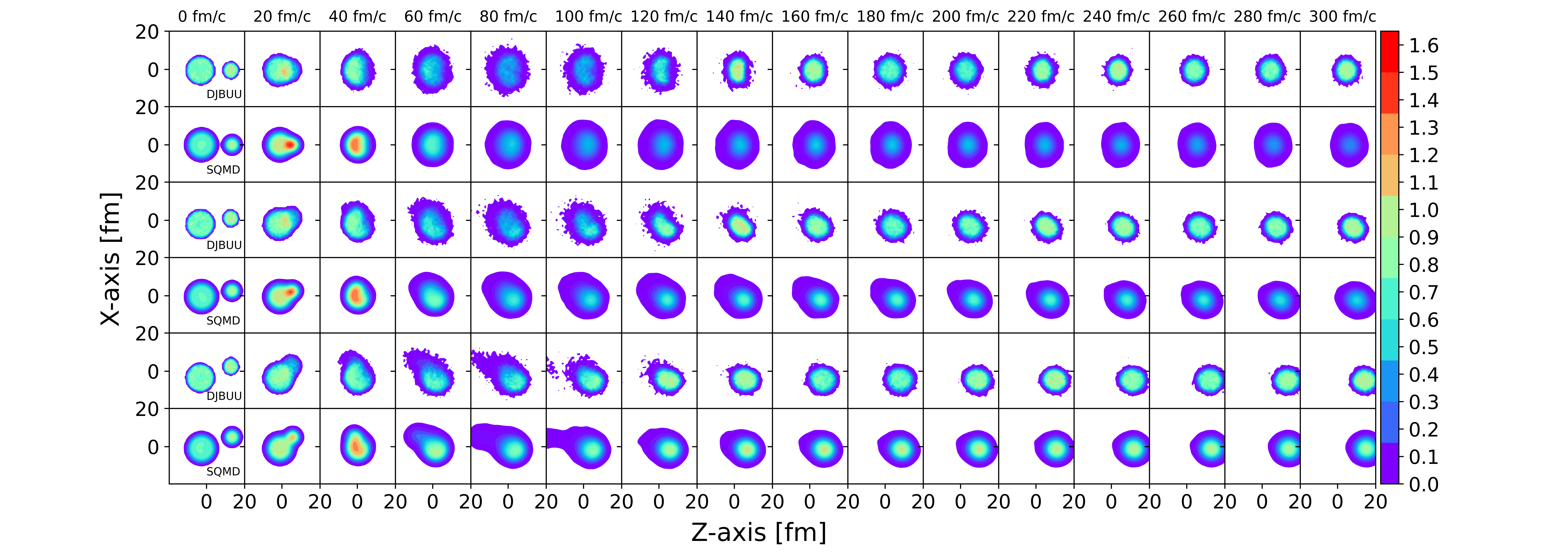}
\includegraphics[width=\linewidth]{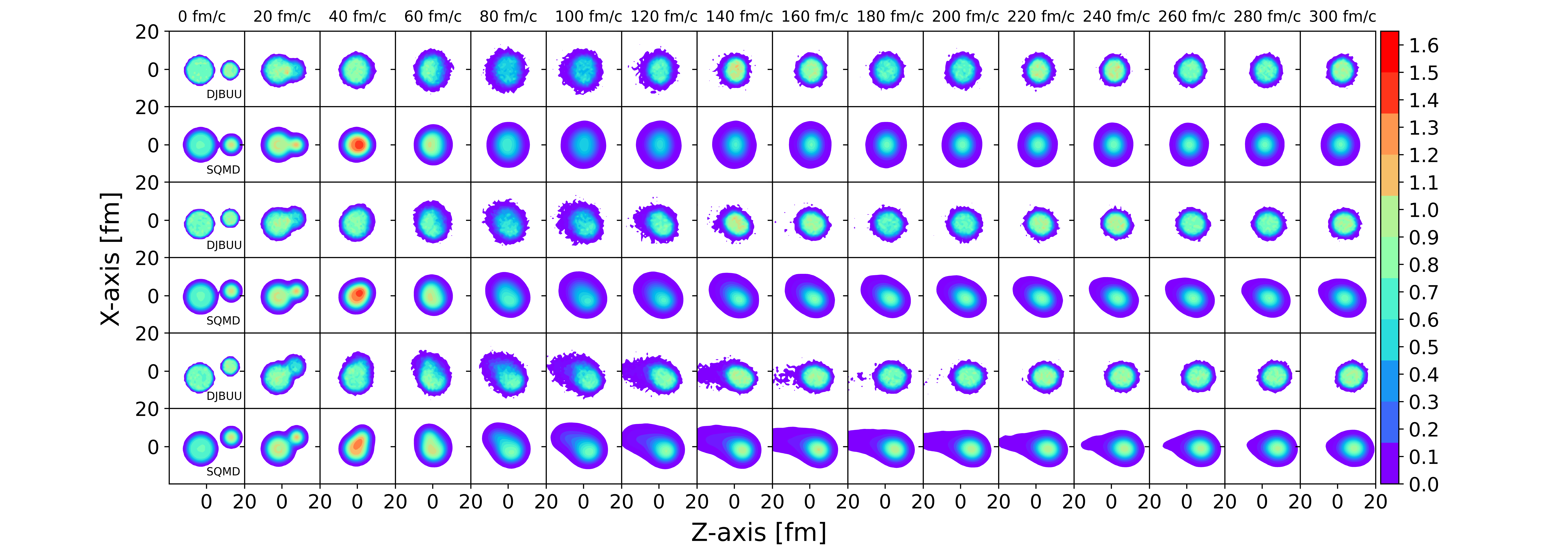}
\caption{Density distribution in the collision plane. For comparison, the results of DJBUU and SQMD are shown alternatively. From top to bottom, the systems are $^{208}$Pb+$^{40}$Ca at $E_{\rm beam}$ =  50 AMeV (top), $^{208}$Pb+$^{40}$Ca at $E_{\rm beam}$ =  100 AMeV (middle) and $^{208}$Pb+$^{48}$Ca at $E_{\rm beam}$ =  50 AMeV (bottom), respectively. In each figure, impact parameters are $b=0$ fm in the first two lows, $b=3$ fm in the middle, and $b=6$ fm in the last two lows,  Central region is plotted up to $20$ fm for the $x$ and $z$ directions. The colors denote the baryon number density in unit of $\rho_0$.}
\label{fig:contour}
\end{figure}

In Fig.~\ref{fig:contour}, the density distributions of three different systems are shown up to $300$ fm/c
simulation time; from the top to bottom, $^{208}$Pb+$^{40}$Ca at $E_{\rm beam}$ = 50 AMeV,
$^{208}$Pb+$^{40}$Ca at $E_{\rm beam}$ = 100 AMeV and $^{208}$Pb+$^{48}$Ca at $E_{\rm beam}$ = 50 AMeV. We
observe that the $^{208}$Pb+$^{48}$Ca and $^{208}$Pb+$^{40}$Ca reactions at $E_{\rm beam}$ = 100 AMeV show
a similar trend and therefore we plot only the case with $^{40}$Ca target at $E_{\rm beam}$ = 100 AMeV.
For  comparison, the results of DJBUU and SQMD are plotted alternatively, and contours for different
impact parameters are displayed, $b$ = 0 fm in the first two rows, $3$ fm in the middle, $6$ fm in the
last. Here, the colors denote baryon number density, and the density is averaged for whole simulation
events.

It can be seen from Fig.~\ref{fig:contour} that when $E_{\rm beam}$ = 50 AMeV, the density has a maximum
value near $t=40$ fm/c. When $E_{\rm beam}$ = 100 AMeV, it has the highest value near $t=30$fm/c, as
expected. A significant difference between DJBUU and SQMD is  the central density. Generally, the BUU-type
models have higher stability than  QMD-type models, which  contributes to the difference. Also, the
difference is partly attributed to the equation of state of nuclear matter adopted in DJBUU and SQMD.

\begin{table}
\centering
\caption{The BFs in DJBUU and SQMD; the BFs from the ten runs of DJBUU 
and the most abundantly produced three BFs from SQMD runs.}
\label{t5}
\begin{tabular}{c|c|c|c|c}
\noalign{\smallskip}\noalign{\smallskip}\hline\hline
Target & \multicolumn{1}{c|}{\begin{tabular}[c]{@{}c@{}}$E_{\rm beam}$\\ {[}AMeV{]}\end{tabular}} & \multicolumn{1}{c|}{\begin{tabular}[c]{@{}c@{}}b \\ {[}fm{]}\end{tabular}}& DJBUU & SQMD \\ 
\hline
\multirow{6}{*}{$^{40}$Ca} & \multirow{3}{*}{50} & 0 &$^{163}_{73}$Ta,\;$^{162}_{73}$Ta,\;$^{164}_{73}$Ta,\;$^{163}_{74}$W & $^{163}_{69}$Tm,\;$^{173}_{74}$W,\;$^{169}_{72}$Hf\\ 
      \cline{3-5}
      & & 3 &$^{163}_{73}$Ta,\;$^{165}_{74}$W,\;$^{164}_{73}$Ta,& $^{169}_{72}$Hf,\;$^{173}_{74}$W,\;$^{172}_{74}$W\\ 
      \cline{3-5}
      & & 6 &$^{167}_{74}$W,\;$^{169}_{75}$Re,\;$^{165}_{73}$Ta,\;$^{168}_{75}$Re & $^{168}_{72}$Hf,\;$^{164}_{70}$Yb,\;$^{169}_{72}$Hf\\ 
      \cline{2-5}
   & \multirow{3}{*}{100} & 0 &$^{123}_{56}$Ba,\;$^{121}_{55}$Cs,\;$^{124}_{57}$La,\;$^{122}_{56}$Ba,\;$^{124}_{56}$Ba & $^{78}_{33}$As,\;$^{114}_{50}$Sn,\;$^{124}_{54}$Xe\\ 
      \cline{3-5}
      & & 3 &$^{130}_{59}$Pr,\;$^{130}_{58}$Ce,\;$^{128}_{57}$La,\;$^{128}_{58}$Ce,\;$^{129}_{58}$Ce,\;$^{127}_{58}$Ce,\;$^{127}_{57}$La & $^{125}_{53}$I,\;$^{128}_{56}$Ba,\;$^{132}_{57}$La\\ 
      \cline{3-5}
      & & 6 &$^{145}_{64}$Gd,$^{144}_{64}$Gd,$^{146}_{65}$Tb,$^{147}_{65}$Tb & $^{151}_{64}$Gd,$^{149}_{63}$Eu,$^{154}_{66}$Dy\\
      \cline{1-5}
\multirow{6}{*}{$^{48}$Ca} & \multirow{3}{*}{50} & 0 &$^{161}_{72}$Hf,$^{162}_{72}$Hf,$^{160}_{71}$Lu,$^{159}_{71}$Lu & $^{167}_{70}$Yb,$^{167}_{71}$Lu,$^{170}_{71}$Lu\\
      \cline{3-5}
      & & 3 &$^{162}_{72}$Hf,$^{164}_{73}$Ta & $^{165}_{70}$Yb,$^{167}_{70}$Yb,$^{167}_{71}$Lu\\
      \cline{3-5}
      & & 6 &$^{164}_{72}$Hf,$^{163}_{72}$Hf,$^{166}_{73}$Ta,$^{165}_{72}$Hf & $^{165}_{69}$Tm,$^{159}_{68}$Er,$^{164}_{69}$Tm\\
      \cline{2-5}			
  & \multirow{3}{*}{100} & 0 &$^{113}_{51}$Sb,$^{115}_{52}$Te,$^{114}_{51}$Sb,$^{116}_{52}$Te,$^{112}_{51}$Sb & $^{58}_{25}$Mn,$^{74}_{32}$Ge,$^{107}_{48}$Pd\\
      \cline{3-5}
      & & 3 &$^{121}_{54}$Xe,$^{122}_{55}$Cs,$^{120}_{54}$Xe,$^{123}_{55}$Cs,$^{121}_{55}$Cs & $^{120}_{52}$Te,$^{106}_{45}$Rh,$^{113}_{48}$Cd\\
      \cline{3-5}
      & & 6 &$^{140}_{62}$Sm,$^{139}_{62}$Sm,$^{138}_{61}$Pm,$^{137}_{61}$Pm,$^{137}_{60}$Nd & $^{147}_{62}$Sm,$^{153}_{64}$Gd,$^{148}_{62}$Sm\\
      \cline{3-5}
      \hline\hline
\end{tabular}
\end{table}

In Table~\ref{t5}, the BFs from DJBUU and SQMD simulations are shown. All the BFs from the ten runs are
presented for DJBUU; while the most abundantly produced three BFs are listed for SQMD. It can be seen from
Table~\ref{t5} that as the beam energy increases from $50$ AMeV to $100$ AMeV, the difference in the
produced BFs between DJBUU and SQMD, especially when $b=0$ fm, becomes significant. This is partly because
the difference in the equations of state adopted in DJBUU and SQMD is more pronounced
at higher beam energies
since the maximum density becomes larger as the bean energy increase. For example, the
symmetry energy used in DJBUU and SQMD can be written as
\begin{align}
	E(\rho,\alpha_I) &= E(\alpha_I =0) + E_{\rm sym} \alpha_I^2 + \cdots,  \label{energy} \\
	E_{\rm sym} &= g_{\rm sym} \left( \frac{\rho}{\rho_0} \right)^\sigma \, , \label{sym_energy}
\end{align}
where $\alpha_I$ is the isospin asymmetry  defined by $\alpha_I=(\rho_n-\rho_p)/(\rho_n+\rho_p)$. In this
study $\sigma=1$ and $\sigma=0.7$ are used for DJBUU~\cite{Liu:2001iz} and SQMD, respectively. We will
discuss the symmetry energy more at the end of this section. Also, since the beam energy ($100$ AMeV)
would be high enough to break nuclei into small pieces, the different definitions of the BF in DJBUU and
SQMD can be more pronounced at the higher beam energy. It can be partly responsible
for the difference in the BF distributions.

Another trend in  Table~\ref{t5} is that  with a fixed beam energy the difference in the produced BFs
between DJBUU and SQMD increases as the impact parameter decreases. This is related to the collision
dynamics and the equation of state. At the central collisions ($b=0$ fm), almost all the nucleons
participate in the collisions. Therefore the bulk of the system evolution depends sensitively 
on the equation of state.
At the peripheral collisions ($b= 6$ fm), however, substantial number of
nucleons do not participate in the collision directly. Hence, the part of the system that 
can feel the equation of state is relatively smaller.

\begin{figure} 
\includegraphics[width=\linewidth]{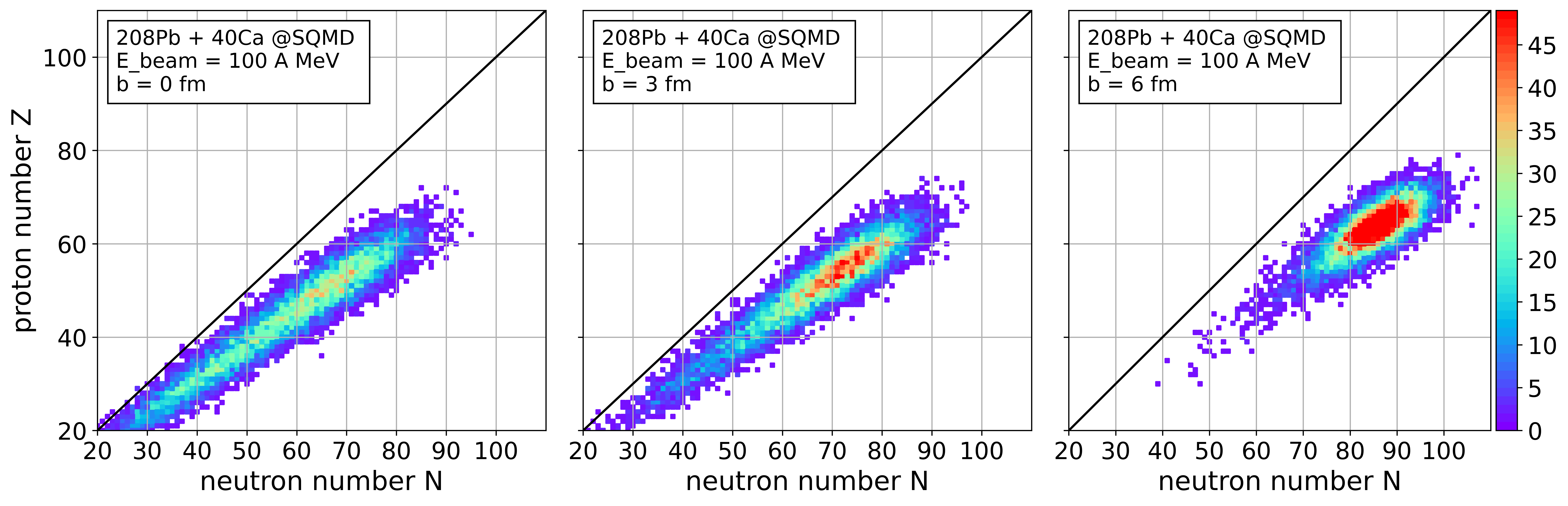}
\caption{Proton and neutron distributions of the BFs in SQMD for $^{208}$Pb+$^{40}$Ca reaction at $E_{\rm beam}$ = 100 AMeV. From left to right, the impact parameters are $b$ = 0 fm (left), 3 fm (center) and 6 fm (right), respectly.}
\label{fig:40_100}
\end{figure}

\begin{figure} 
\includegraphics[width=\linewidth]{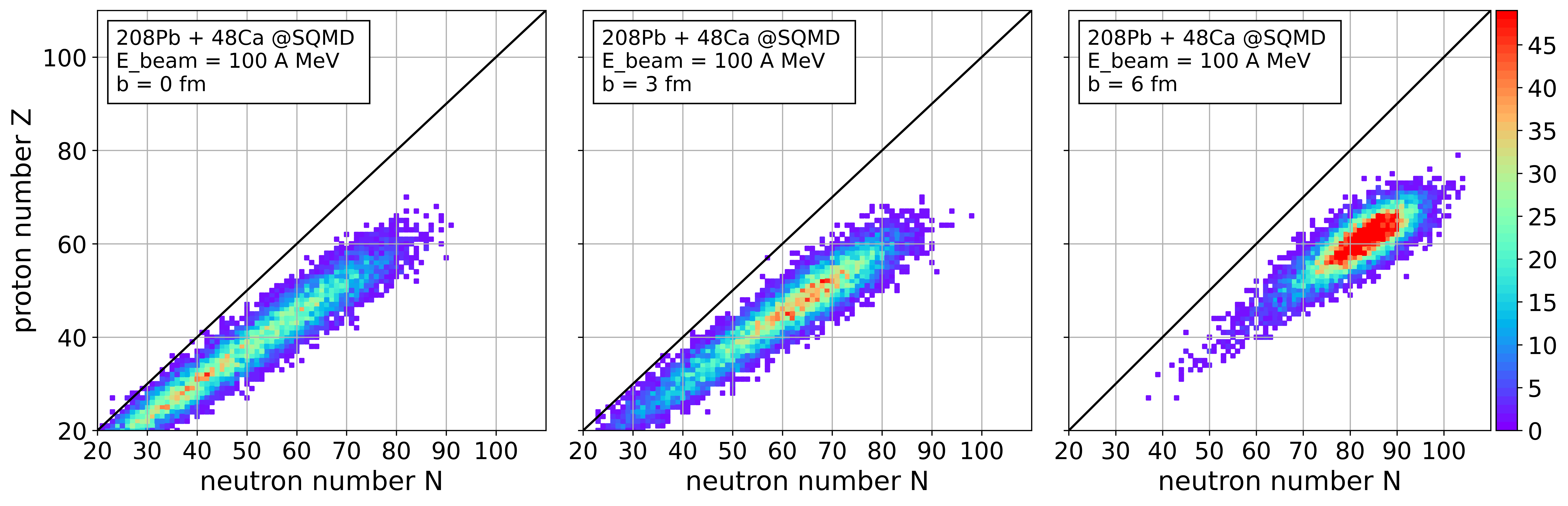}
\caption{The same as Fig.~\ref{fig:40_100} for $^{208}$Pb+$^{48}$Ca reaction.
}
\label{fig:48_100}
\end{figure}

\begin{figure} 
\includegraphics[width=0.45\linewidth]{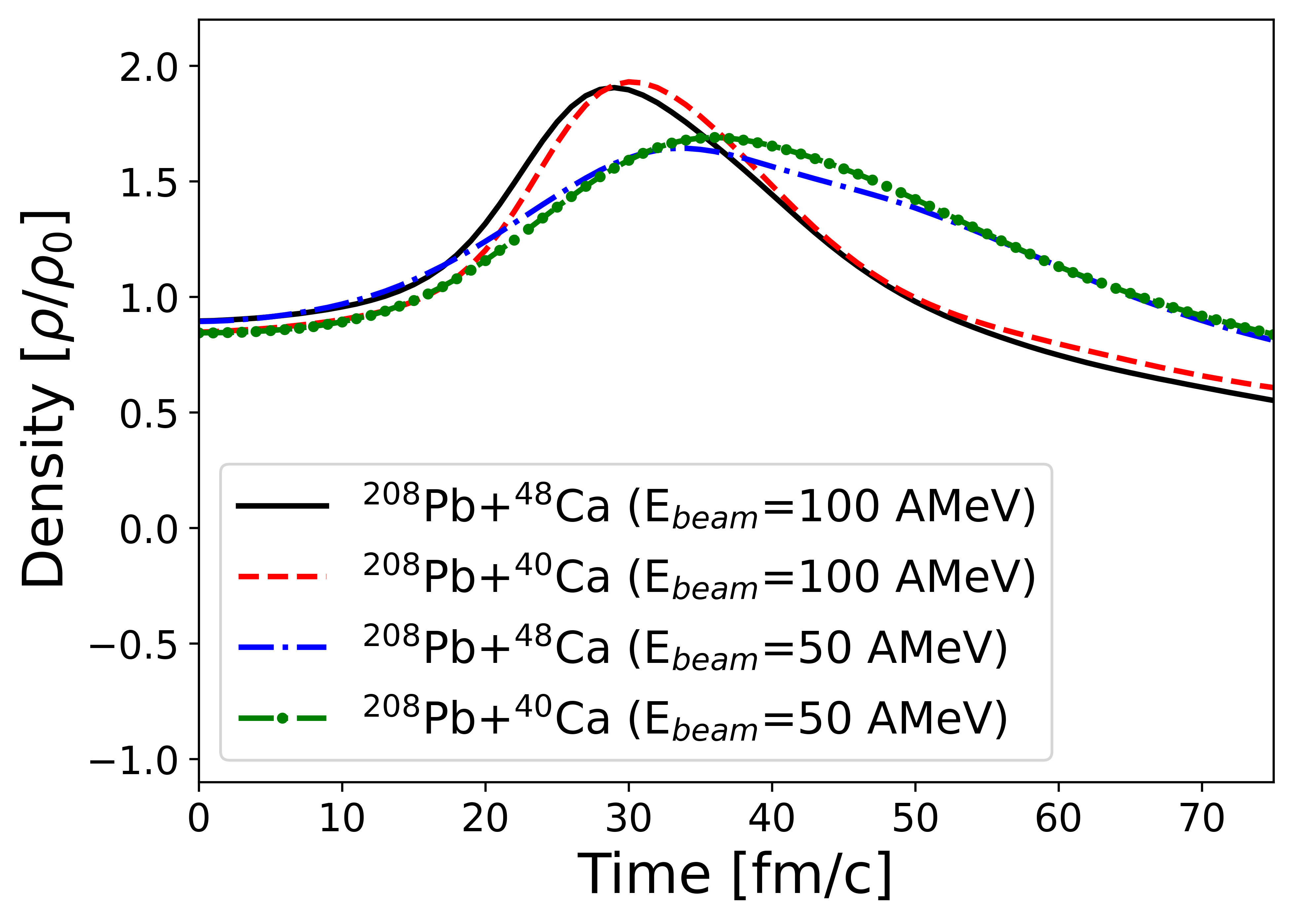}
\includegraphics[width=0.45\linewidth]{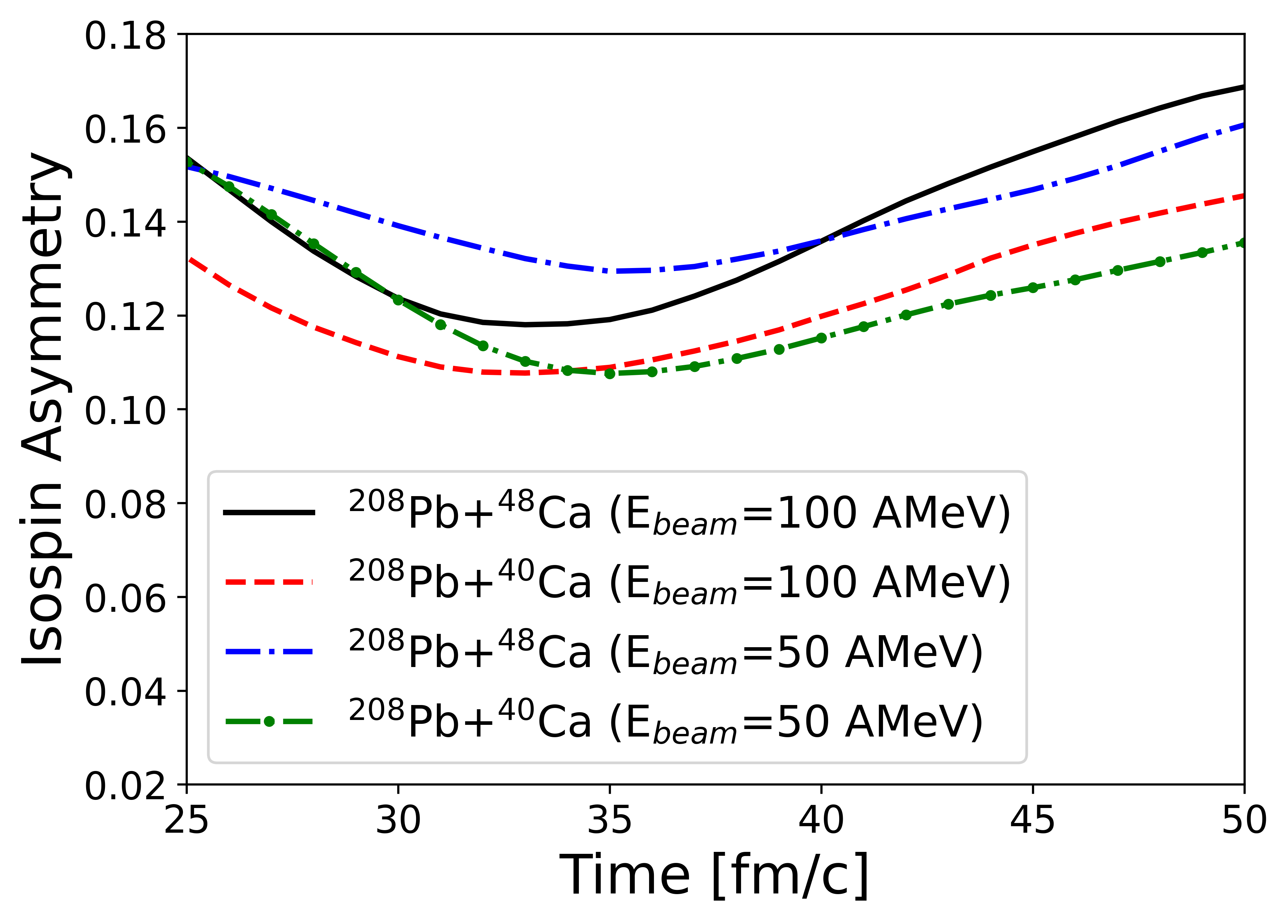}
\caption{Baryon number density(left) and isospin asymmetry(right) as a function of the simulation time at the collision center for the central collsions ($b$ = 0 fm). Black and red lines are for $^{208}$Pb+$^{48}$Ca and $^{208}$Pb+$^{40}$Ca reactions with $E_{\rm beam} = 100$ AMeV, and blue and green lines are for $^{208}$Pb+$^{48}$Ca and $^{208}$Pb+$^{40}$Ca reactions with $E_{\rm beam} = 50$ AMeV.}
\label{density}
\end{figure}

To investigate some features of the BFs produced in SQMD, we show the proton and neutron distributions of
the BFs in SQMD for $^{208}$Pb+$^{40}$Ca and $^{208}$Pb+$^{48}$Ca reactions at $E_{\rm beam}$ =  100 AMeV
in Fig.~\ref{fig:40_100} and Fig.~\ref{fig:48_100}, respectively. From the left to the right, we plot the
distributions with the impact parameters 0, 3 and 6 fm. The colors denotes  the total number of the
produced BF in ten-thousand simulation events. A trend we can observe in  Fig.~\ref{fig:40_100} and
Fig.~\ref{fig:48_100} is that for the central collisions ($b$ = 0 fm), the ranges of proton and neutron
numbers are relatively wide, while those become narrow for the peripheral collisions ($b$ = 6 fm).
One of the interesting results is that, even though the neutron number of the system is increased and the
bombarding energy in the center of mass frame is decreased from the reaction $^{208}$Pb+$^{40}$Ca to
$^{208}$Pb+$^{48}$Ca, the proton and neutron numbers of the BFs for $^{208}$Pb+$^{48}$Ca reactions are
smaller than those for $^{208}$Pb+$^{40}$Ca reactions at $E_{\rm beam}$ =  100 AMeV. One way to understand
this phenomenon is the role of the nuclear symmetry energy. This effect is more significant in
the $^{208}$Pb+$^{48}$Ca reactions than $^{208}$Pb+$^{40}$Ca reactions. In a neutron-rich matter, the
symmetry energy pushes out the neutrons and so disturbs the formation of large fragments.
To check the effect of the symmetry energy, we calculate the baryon density and the isospin asymmetry
$\alpha_I$ as a function of time at the collision center and show them in Fig.~\ref{density}. Note that as
in Eq. (\ref{energy}), in the expression of the energy of a system, the symmetry energy $E_{\rm sym}$
comes with the isospin asymmetry as $E_{\rm sym}\alpha_I^2$.
As can be
seen in the right panel of Fig.~\ref{density}, the isospin asymmetry in the
$^{208}$Pb+$^{48}$Ca reaction is overall bigger that the one in the $^{208}$Pb+$^{40}$Ca reaction.
Therefore, we expect that the role of the symmetry energy is more pronounced in the $^{208}$Pb+$^{48}$Ca
reaction, resulting in smaller BFs compared to the $^{208}$Pb+$^{40}$Ca reaction.

\section{Summary}

In this work, we have compared two transport model codes DJBUU and SQMD. Both models have been
developed mainly for RAON physics where one of the main goals is to produce rare isotope.
To this end, we have investigated the fragment productions in
$^{208}$Pb+$^{40}$Ca and $^{208}$Pb+$^{48}$Ca reactions at $50$ and $100$ AMeV with the impact parameters
$b=0,~3,~6$ fm, focusing on primary fragments from DJBUU and SQMD. We observed that overall the two models
produce similar fragments. For example, though the fragments with $^{48}$Ca are a bit smaller than those
with $^{40}$Ca, the produced fragments from the both models are in the range of ${\rm A}= 162\sim 168$
with ${\rm Z}=70\sim 74$, where A is the atomic number and Z is the charge number. We found a noticeable
difference between DJBUU and SQMD at $E_{\rm beam}=100$ AMeV with $b$ = 0 fm, which can be attributed to
the difference in the equation of state used in the models and also  to the difference in stability
inherent to  the BUU-type and  QMD-type models. After some improvements such as including a surface term
(time-dependent width of the nucleons) in DJBUU (SQMD), DJBUU and SQMD will play important roles in
studying heavy ion reactions and related physics.


\begin{acknowledgments}
The work of D.I.K. and C.H.L. was supported by  National Research Foundation of Korea (NRF) grants funded by the
Korea government (Ministry of Science and ICT and Ministry of Education)  (No. 2016R1A5A1013277 and No.
2018R1D1A1B07048599). The work of K.K.~and Y.K.~was supported partly by the Rare Isotope Science Project of
Institute for Basic Science funded by Ministry of Science, ICT and Future Planning, and NRF of Korea
(2013M7A1A1075764). This work was supported by the National Supercomputing Center with supercomputing
resources including technical support(KSC-2021-CRE-0477 and KSC-2021-RND-0076).
S.J.~is supported in part by the Natural Sciences and Engineering Research Council of Canada.
\end{acknowledgments}



\begin{references}

%
\bibitem{Tshoo:2013voa}
K.~Tshoo, Y.~K.~Kim, Y.~K.~Kwon, H.~J.~Woo, G.~D.~Kim, Y.~J.~Kim, B.~H.~Kang, S.~J.~Park, Y.~H.~Park and J.~W.~Yoon, \textit{et al.}
Nucl. Instrum. Meth. B \textbf{317}, 242 (2013).

%
\bibitem{Jeong:2018qvi}
S.~Jeong, P.~Papakonstantinou, H.~Ishiyama and Y.~Kim,
J. Korean Phys. Soc. \textbf{73}, 516 (2018).



\bibitem{DJBUU2016}
Myungkuk Kim, Chang-Hwan Lee, Youngman Kim and Sangyong Jeon,
New Physics: Sae Mulli {\bf 66}, 1563 (2016).

\bibitem{Kim:2020sjy}
M.~Kim, S.~Jeon, Y.~M.~Kim, Y.~Kim and C.~H.~Lee,
Phys. Rev. C \textbf{101}, 064614 (2020).


%
\bibitem{Kim:2017hmt}
K.~Kim, Y.~Kim and K.~S.~Lee,
J. Korean Phys. Soc. \textbf{71}, 628 (2017).



%
\bibitem{Mocko:2008rj}
M.~Mocko, M.~B.~Tsang, D.~Lacroix, A.~Ono, P.~Danielewicz, W.~G.~Lynch and R.~J.~Charity,
Phys. Rev. C \textbf{78}, 024612 (2008).

\bibitem{Gaitanos:2007mm}
T.~Gaitanos, H.~Lenske and U.~Mosel,
Phys. Lett. B \textbf{663}, 197 (2008).


\bibitem{Napolitani:2009pc}
P.~Napolitani, M.~Colonna, F.~Gulminelli, E.~Galichet, S.~Piantelli, G.~Verde and E.~Vient,
Phys. Rev. C \textbf{81}, 044619 (2010).

\bibitem{Liu:2014poa}
X.~Liu, W.~Lin, R.~Wada, M.~Huang, S.~Zhang, P.~Ren, Z.~Chen, J.~Wang, G.~Q.~Xiao and R.~Han, \textit{et al.}
Nucl. Phys. A \textbf{933}, 290 (2015).

%
\bibitem{TMEP:2022xjg}
H.~Wolter \textit{et al.} [TMEP],
Prog. Part. Nucl. Phys. \textbf{125}, 103962 (2022).


\bibitem{Motohiro:2015taa}
Y.~Motohiro, Y.~Kim and M.~Harada,
Phys. Rev. C \textbf{92},  025201 (2015).

\bibitem{Shin:2018axs}
I.~J.~Shin, W.~G.~Paeng, M.~Harada and Y.~Kim,
[arXiv:1805.03402 [nucl-th]].

\bibitem{Takeda:2017mrm}
Y.~Takeda, Y.~Kim and M.~Harada,
Phys. Rev. C \textbf{97} 065202 (2018).

\bibitem{Danielewicz:1992mi}
P.~Danielewicz and Q.~b.~Pan,
Phys. Rev. C \textbf{46}, 2002-2011 (1992).

\bibitem{Aichelin:1991xy}
J.~Aichelin,
Phys. Rept. \textbf{202}, 233-360 (1991).

\bibitem{Papa:2002xqa}
M.~Papa, T.~Maruyama and A.~Bonasera,
{\it ``Constrained Molecular Dynamics Approach to Fermionic Systems''},
Nuclear Physics at Border Lines, pp. 253-256 (World Scientific, 2002);
doi:10.1142/9789812778321\_0041

\bibitem{Liu:2001iz}
B.~Liu, V.~Greco, V.~Baran, M.~Colonna and M.~Di Toro,
Phys. Rev. C \textbf{65}, 045201 (2002).

\end{references}
\end{document}